\documentclass[letter,twocolumn]{jpsj3}
\usepackage{txfonts}

\usepackage{physics}
\usepackage{graphicx,color}

\title{Evolution and Instability of Bogoliubov Fermi Surfaces under Zeeman Field}

\author{Tatsuaki Mori$^1$, Hiroshi Watanabe$^2$, and Hiroaki Ikeda$^1$}
\inst{$^1$Department of Physical Science, Ritsumeikan University, Kusatsu 525-8577, Japan \\
$^2$College of Industrial Technology, Nihon University, Narashino 275-8576, Japan} 

\abst{{We theoretically investigate the evolution and instability of the Bogoliubov Fermi surface (BFS) in the spherical $j=3/2$ model under a Zeeman field. The applied field induces a pronounced expansion in the BFS with $j_z = \pm 3/2$ component.} Such behavior can be detected by spectroscopic techniques such as angle-resolved photoemission spectroscopy (ARPES). {Interestingly, the Pauli susceptibility exhibits behavior that appears discontinuous just below the transition temperature at zero field, even though it is a second-order transition. This is due to spontaneous magnetization.} Furthermore, the analysis of the bogolon correlations in the superconducting phase suggests the possibility of the chiral $p$- or $f$-wave bogolon pairing instabilities rather than {the Pomeranchuk instability.} These chiral states coexist with the chiral $d$-wave superconducting state, spontaneously break inversion symmetry, and lead to the disappearance of the torus-shaped BFS structure.
}

\newcommand{\onlinecite}[1]{[\nocite{#1}\citenum{#1}]}

\begin{document}
\maketitle
\thispagestyle{plain}
\pagestyle{plain}

The gap structure is essential for understanding the mechanisms of superconductivity. Since the discovery of high-temperature superconductors in cuprates, significant attention has been devoted to anisotropic gap structures, such as line nodes. Recently, it has been shown that Bogoliubov quasiparticles in certain superconductors can form a stable Fermi surface (FS), known as the Bogoliubov Fermi Surface (BFS), which has attracted considerable interest~\cite{VOLOVIK1989282,VOLOVIK1989JETP,PhysRevLett.118.127001,PhysRevB.100.224505,PhysRevB.98.224509,PhysRevB.105.134507,PhysRevB.107.L220501,PhysRevX.8.011029,PhysRevResearch.2.033013,PhysRevB.101.184503,PhysRevB.104.094529,PhysRevLett.127.187003,PhysRevB.107.214513,PhysRevB.108.014502,PhysRevB.99.134513,PhysRevB.108.224503,Babkin_2024}. 
The emergence of a BFS is closely related to a pseudo-magnetic field arising from interband Cooper pairing, which is driven by the time-reversal symmetry breaking (TRSB). 
Several unconventional superconductors have been proposed to exhibit BFS, further intensifying interest in this phenomenon. Notably, superconductors such as UPt$_3$ \cite{PhysRevLett.68.117,PhysRevLett.71.1466,doi:10.1126/science.1248552}, URu$_2$Si$_2$ \cite{PhysRevB.91.140506}, and Sr$_2$RuO$_4$ \cite{arXiv:condmat9808159,PhysRevLett.97.167002,Kittaka_2018,PhysRevResearch.2.032023} have long been regarded as multiband superconductors exhibiting TRSB, making the formation of a BFS within these systems plausible.
Moreover, recent experimental studies on Fe(Se,S) using techniques such as specific heat, thermal conductivity \cite{doi:10.1073/pnas.1717331115}, STM \cite{doi:10.1126/sciadv.aar6419}, NMR \cite{Yu_2023}, $\mu$SR \cite{doi:10.1073/pnas.2208276120}, and laser angle-resolved photoemission spectroscopy (ARPES) \cite{doi.org/10.21203/rs.3.rs-2224728/v1} have revealed a finite density of states (DOS) at the Fermi level, alongside evidence of TRSB, suggesting the presence of BFS. This pairing state has been referred to as ``ultra-nodal pairing state'' \cite{Setty_2020,PhysRevB.102.064504,PhysRevB.108.224506,Islam2024-ig,PhysRevB.110.L020503}.

Given this context, a deeper understanding of BFS properties is essential to link theoretical predictions with experimental findings in real materials. Theoretically, various aspects of BFS have already been explored, including how it could be experimentally observed \cite{PhysRevB.101.024505}, the impurity effects \cite{PhysRevLett.127.257002,PhysRevB.104.094518,PhysRevB.109.094502}, the formation of odd-frequency Cooper pairs~\cite{Kim_2021,PhysRevResearch.3.033255}, optical responses \cite{Ahn_2021}, and {transport properties}~\cite{PhysRevB.110.104515,Pal_2024,pal2024identifyinginflatedfermisurfaces}.
Furthermore, it has been discussed that, similar to the Fermi surface in the normal state, appropriate interactions could destabilize the BFS, potentially inducing new phase transitions. Several studies have suggested that the BFS could trigger Pomeranchuk or Cooper instabilities, resulting in a complex superconducting phase diagram \cite{PhysRevB.102.024505,PhysRevB.102.020501,PhysRevB.103.024521,PhysRevB.103.144517}.
Despite these advancements, much of the research has been limited to zero-temperature or zero-magnetic field conditions, leaving the behavior of BFS under finite temperature and magnetic fields largely unexplored. This paper aims to address this gap by investigating the properties of BFS in the presence of the Zeeman field, thereby deepening our understanding of BFS. Specifically, we identify novel orders of bogolons as a function of temperature and Zeeman field in a $j=3/2$ electron model. The study is framed in the context of spin-quintet superconductivity, widely classified as spin-singlet, and also provides valuable insights into materials such as Fe(Se,S).

\begin{figure*}[ht]
\centering
\includegraphics[width=0.9\textwidth]{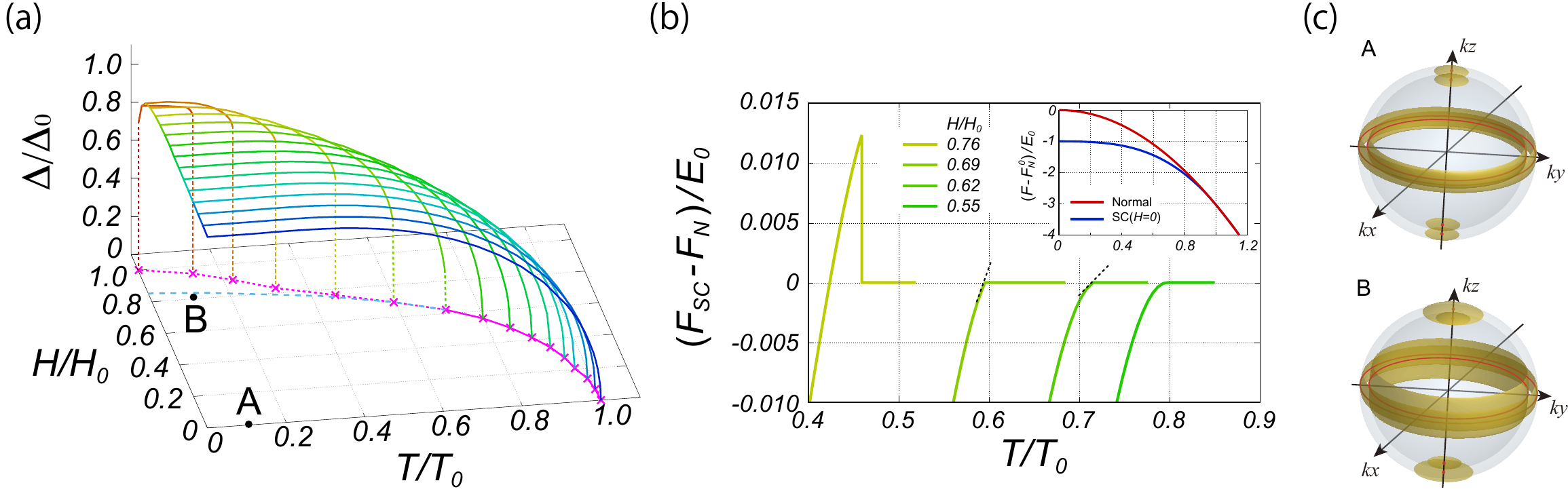}
\caption{{(a) $T$ dependence of the gap ratio $\Delta/\Delta_0$ for various Zeeman fields and $H-T$ phase diagram. Here, $\Delta_0$ denotes the gap amplitude at $T=0$ and $H=0$. $T$ and $H$ are normalized by the zero-field transition temperature $T_0$ and the highest critical field $H_0=\gamma T_0$ ($\gamma\simeq1.65$), respectively. Dashed lines for $\Delta$ indicate discontinuities due to a first-order transition. 
In the $H-T$ phase diagram, the pink line highlights the $T$ dependence of the critical field as a guide to the eye, while the dashed line indicates a first-order transition. The light blue broken line represents a thermodynamically stable critical field.
(b) {$T$ dependence of} normalized free energy difference {for several $H$.} For $H/H_0=0.55$, the slope changes smoothly at the transition point, while for $H/H_0=0.62$ and $0.69$, a kink appears at the transition. (Black dotted lines represent the slope just below the transition temperature.)
For $H/H_0=0.76$, $F_{SC}-F_N>0$, indicating that the superconducting state is metastable.
{The inset depicts} the free energy normalized by the condensation energy $E_0=\Delta^2/V$. Here, $F_N^0$ denotes the normal-state free energy $F_N(T=0)$ at $T=0$.
{(c) 3D cartoons in panels A and B illustrate the gap structure (BFS) corresponding to points A and B in the $H-T$ phased diagram.
The transparent sphere illustrates the normal-state FS at $k_F=\sqrt{\mu/(\alpha-5\beta/4\pm\beta)}$.~\cite{PhysRevB.100.224505}
The red points along $k_z$-axis and the red horizontal line in the $k_z=0$ plane represent the symmetry-protected nodes characteristic of the $d_{xz}+id_{yz}$-wave pairing state.The gold-colored surface indicates the BFSs.}}}
\end{figure*}

{\it Formulation}---
As a simple model exhibiting a BFS, we consider the spherical $j=3/2$ model introduced by Agterberg et al.~\cite{PhysRevLett.118.127001,PhysRevB.98.224509,PhysRevLett.116.177001,PhysRev.97.869}, to elucidate the evolution and instability of the BFS under the Zeeman field.
The Hamiltonian~{\cite{Zeeman}} is given by
\begin{align}
    \mathcal{H}&=\sum_{\textbf{k}}\psi_{\textbf{k}}^{\dagger}
    \left[(\alpha\textbf{k}^2-\mu)\hat{1}+\beta(\textbf{k}\cdot\hat{\textbf{J}})^2-g\mu_B\hat{J}_z H\right]
    \psi_{\textbf{k}} \nonumber \\
    &+\sum_{\textbf{k}}\left(\psi_{\textbf k}^\dag\hat{\Delta}_{\textbf k}\psi_{-\textbf k}^{\dag T}+{\rm H.c.}\right),
\end{align}
where $\psi_{\textbf{k}}=(c_{\textbf{k},\frac{3}{2}},c_{\textbf{k},\frac{1}{2}},c_{\textbf{k},-\frac{1}{2}},c_{\textbf{k},-\frac{3}{2}})^T$
is the spin-$3/2$ spinor consisting of electron annihilation operators $c_{\textbf{k},j_z}$ with wave vector $\textbf k$ and the $z$ component of angular momentum $j_z$. 
$\hat{\Delta}_{\textbf{k}}$ denotes the $4\times4$ matrix of superconducting gap functions, $\hat{\textbf J}$ the angular momentum matrix for $j=3/2$, and $\hat{1}$ the $4\times4$ identity matrix. 
$\alpha$, which gives the energy scale of the system, is set to $1$ hereafter.
$\beta$ and $H$ are the symmetric spin-orbit coupling and the Zeeman field parallel to the $z$ axis, respectively. 
The $g$ factor and the Bohr's magneton $\mu_B$ are taken as $1$ unless otherwise noted.
Electron filling is determined so that $\mu=1$ at $T=0$. 
Through the Bogoliubov transformation, the Hamiltonian is diagonalized as follows,
\begin{equation}
    \mathcal{H}=\sum_{\textbf{k}n} E_{\textbf{k}n}\alpha^\dag_{\textbf{k}n}\alpha_{\textbf{k}n},
\end{equation}
where $\alpha_{\textbf{k}n}$ is the annihilation operator of a Bogoliubov quasiparticle (bogolon) and $E_{\textbf{k}n}$ is the excitation energy of the bogolon.
Here, we consider an on-site pairing interaction $V_0$ and focus on the following local quintet state with TRSB \cite{PhysRevLett.118.127001,PhysRevB.98.224509,PhysRevB.100.224505},
\begin{equation}
    \label{eq:delta}
    \hat{\Delta}_{\textbf k}=\Delta\,(\hat{\eta}_{xz}+i\,\hat{\eta}_{yz})\equiv \Delta\, \hat{\eta}_+,
\end{equation}
where $\hat{\eta}_{xz}=(\hat{J}_x \hat{J}_z+\hat{J}_z \hat{J}_x) \hat{U}_T$, 
$\hat{\eta}_{yz}=(\hat{J}_y \hat{J}_z+\hat{J}_z \hat{J}_y) \hat{U}_T$, and
$\hat{U}_T$ is an antisymmetric matrix corresponding to the time-reversal operation multiplied by the complex conjugate.
This local quintet pairing state with TRSB induces the chiral $d$-wave pair amplitude \cite{PhysRevLett.116.177001}, giving rise to point nodes along the $k_z$-axis and line nodes on the $k_z=0$ plane. Additionally, the spontaneously induced magnetic field expands these nodes, forming BFSs \cite{PhysRevLett.118.127001,PhysRevB.98.224509,PhysRevB.100.224505}.
The gap amplitude $\Delta$ is determined by solving the following BCS gap equation,
\begin{align}
    \label{eq:gap}
    \Delta={V_0\over 8}{1\over 4\pi}\! \int \! d\Omega \int_{k_F-\delta k}^{k_F+\delta k} \hspace{-6mm} k^2 dk \,
    \Tr[\hat{\eta}_+^\dag \ev{\psi_{-\textbf{k}} \psi_{\textbf{k}}^T}]
\end{align}
The integration over $k$ is performed with a step size of $1/50000$ within $\delta k=0.06$ around the Fermi wavenumber $k_F$ at each band.
The integration for the polar angles is computed numerically using $2500$ meshes, while the azimuthal angles are treated analytically.
Hereafter, we set $\beta=-0.1$ and $V_0=2.16$, in which a TRSB pairing state with BFSs is expected to emerge \cite{PhysRevB.100.224505}. 
In our case, a local TRSB quintet state with BFSs [Eq.(\ref{eq:delta})] is obtained below the {zero-field} transition temperature $T_0\simeq 0.00876$.
In the following, we examine the effect of the Zeeman field based on this case.

{\it Phase diagram and discontinuous transition}---
First, we consider the temperature dependence of the gap amplitude $\Delta$ and the $H-T$ phase diagram in Fig.\,1(a).
At zero field, the temperature dependence of $\Delta$ shows conventional BCS-like behavior, while $\Delta$ at $T=0$ remains nearly constant, except in the vicinity of the {highest} critical field $H_0=\gamma T_0$ with $\gamma\simeq 1.65$. 
Under high fields, $\Delta$ exhibits a discontinuity at the transition point for $H/H_0\gtrsim 0.65$, indicative of the first-order transition.
This is described by the dashed pink line in the $H-T$ phase diagram, while the solid line represents a second-order transition.
Such behavior is consistent with previous studies based on the Ginzburg-Landau theory \cite{SARMA19631029,10.1143/PTP.31.945,sato2024discontinuoustransitionsuperconductingphase}.

\begin{figure}[tb]
\centering
\includegraphics[width=0.4\textwidth]{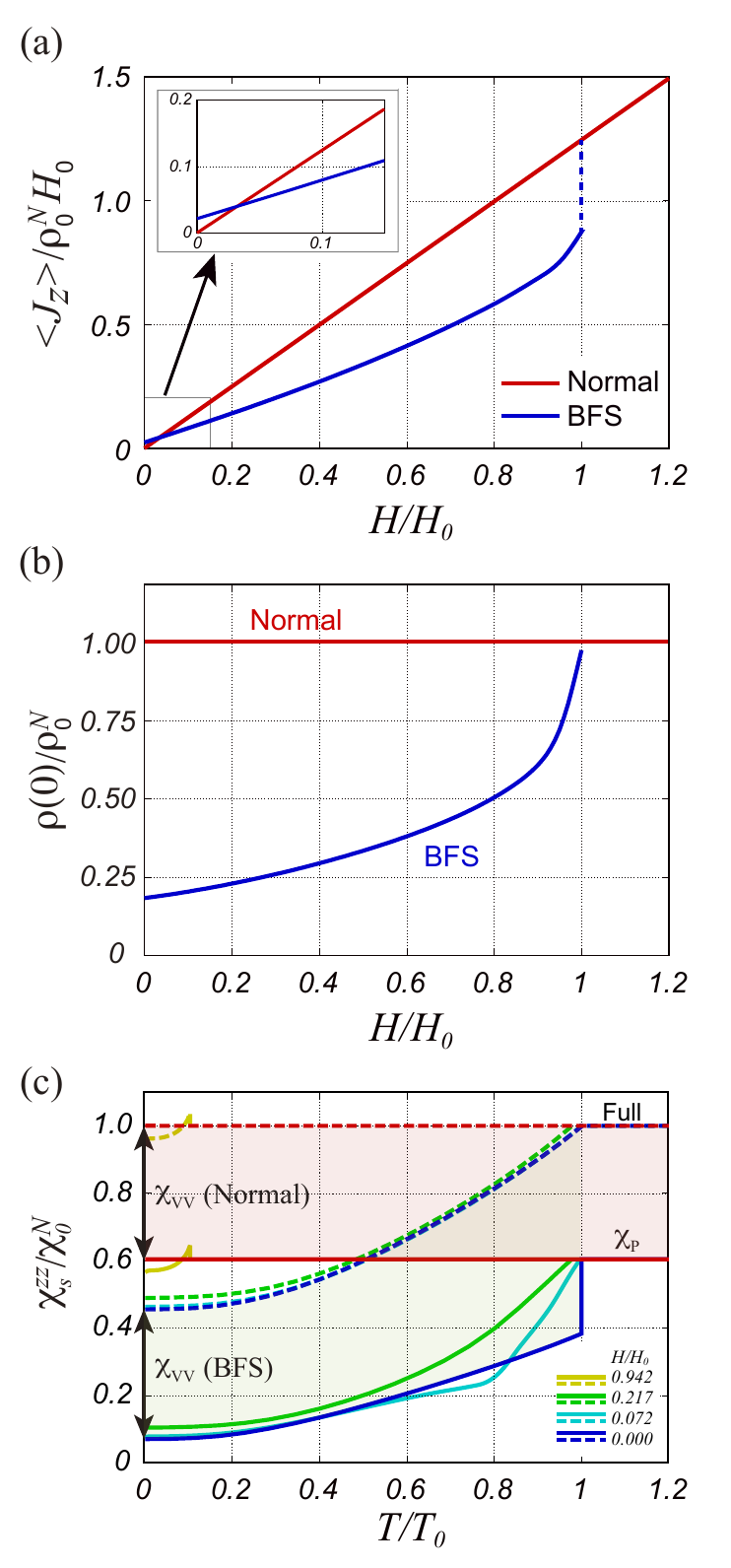}
\caption{{(a) $H$ dependence of magnetization $\ev{J_z}$ at $T=0$, normalized by $\rho_0^N H_0$. 
The inset shows an enlarged view at low fields, where a finite spontaneous magnetization is observed in the spin-quintet state (blue line), even at $H=0$. 
(b) The DOS at the Fermi level $\rho(0)$, normalized by the normal-state value $\rho_0^N$. 
(c) $T$ dependence of $\chi_s^{zz}/\chi_0^N$ (dashed lines) and the Pauli term $\chi_P$ (solid lines) for several $H$. 
The transparent hatching represents the van Vleck contribution in the normal state (red) and for $H=0$ (green).
For $H=0$, $\chi_P$ (blue solid line) exhibits a discontinuous-like behavior just below $T_0$, while such behavior is not observed in $\chi_s^{zz}/\chi_0^N$ (dashed lines). 
The peculiar behavior disappears for a small applied field.
Under high fields (gold-colored line), $\chi_s^{zz}/\chi_0^N$ exhibits a discontinuous jump, reflecting the first-order transition.}}
\end{figure}

{
Next, we examine the thermodynamic stability of superconductivity{\cite{PhysRevB.100.224505}}. {The inset of Fig.\,1(b)} shows the free energy as a function of temperature.
As expected, the free energy decreases monotonically with temperature. The red line corresponds to the normal-state free energy $F_N$.
At $H=0$, the free energy of the superconducting state $F_{SC}$ is significantly lower than $F_N$. The energy difference {from $F_N$} decreases with the applied field.
{The main panel of Fig.\,1(b)} provides an enlarged view of the temperature dependence of $F_{SC}-F_{N}$ around the region where the order of the transition changes.
For $H/H_0\simeq 0.55$, the condensation energy smoothly approaches zero, while for $H/H_0\simeq 0.62$ and $0.69$, a distinct kink structure appears at the transition point. Furthermore, for $H/H_0\simeq 0.76$, a discontinuous jump is observed. This indicates that the superconducting state has a higher energy.
The observed kink structures and discontinuous jumps provide clear evidence for a first-order transition.
In particular, the discontinuous jumps suggest that the superconducting state is metastable.
The thermodynamically stable critical field is shown in {Fig.\,1(a)} with a light-blue dashed line.
}

{\it Evolution of BFS}---
Panels A and B in {Fig.\,1(c)} present a 3D schematic view of a typical BFS. The gold-colored surface indicates the BFSs. 
The BFSs near the $k_z$-axis are inflated point nodes, whereas the torus-shaped BFSs near the $k_z=0$ plane are inflated horizontal line nodes. 
Under high fields, bands dominated by $j_z=\pm 3/2$ components undergo significant shifts, particularly near the $k_z=0$ plane of inside FS and along the $k_z$-axis of the outside FS~{\cite{SM1}}.
This shift results in a pronounced energy separation at high fields, leading to an enlargement of the BFSs~\cite{mori_ICM2024}.
In general, FSs with large spin splitting induced by the Zeeman field tend to generate large BFSs. The evolution of BFSs under external fields can be experimentally observed using ARPES.

\begin{figure*}[tb]
\includegraphics[width=\textwidth]{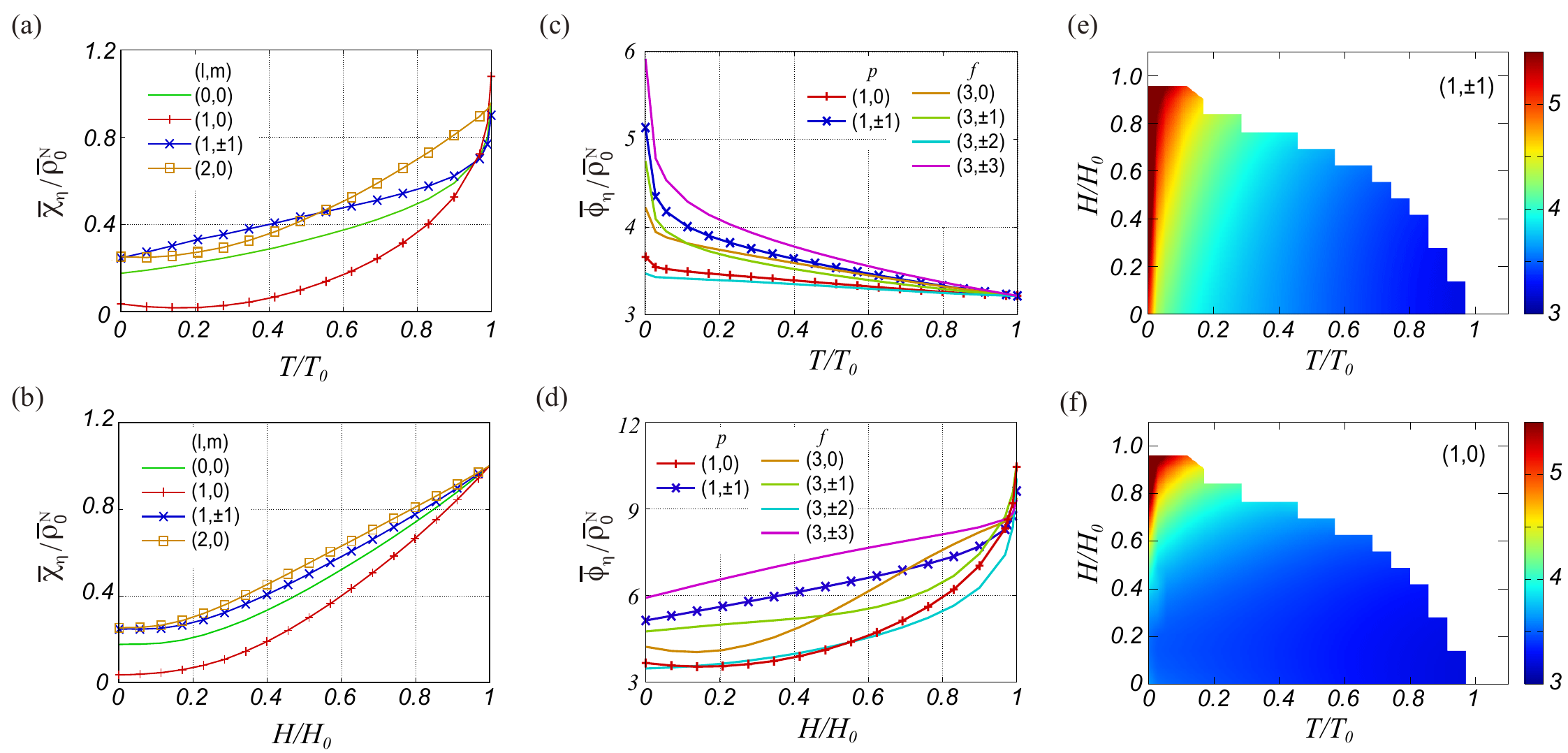}
\caption{Bogolon correlations $\bar\chi_\eta$ as a function of temperature (a) and Zeeman field (b). Those values are normalized by $\bar\rho_0^N=\rho_0^N/4\pi$. Bogolon pair correlations $\bar\phi_\eta$ as a function of temperature (c) and Zeeman field (d). The contour map of (e) $\bar\phi_{x/y}$ and (f) $\bar\phi_z$.}
\end{figure*}


{\it Magnetization and spin susceptibility}---
Figure\,2(a) and (b) show the field dependence of magnetization at $T=0$ and the density of states (DOS) at the Fermi level, $\rho(0)$, respectively.
The magnetization is given by
\begin{equation}
    \ev{J_z}=\sum_{\textbf{k}}\sum_{n} \mel{\textbf{k}n}{\hat{J}_z}{\textbf{k}n}  f(E_{\textbf{k}n}).
\end{equation}
{We can confirm that a small but finite magnetization is observed even at $H=0$ in the superconducting state (inset of Fig. 2(a)), and correspondingly,
$\rho(0)$ remains finite (Fig. 2(b)). These are hallmarks of the chiral state with BFSs.}
Under an applied field, the magnetization increases nearly linearly with $H$ and reaches approximately $0.7$ times the normal-state value at $H=H_0$.
{At this point, $\rho(0)$ almost recovers its normal-state value, suggesting that the deviation of $\ev{J_z}$ from the normal-state value arises primarily from changes in the wave function.}

Figure\,2(c) illustrates the temperature dependence of the spin susceptibility $\chi_s^{zz}$, expressed as
\begin{subequations}
\begin{align}
    \chi_s^{zz}  &=\chi_{\rm P}+\chi_{\rm VV}\,, \\
    \chi_{\rm P} &=\sum_{\textbf{k}}\sum_{n} |\mel{\textbf{k}n}{{\hat J_z}}{\textbf{k}n}|^2 \left(-{\partial f(E_{kn}) \over \partial E_{kn}} \right)\,, \\
    \chi_{\rm VV} &=\sum_{\textbf{k}}\sum_{n\ne n'} |\mel{\textbf{k}n}{{\hat J_z}}{\textbf{k}n'}|^2 \,{f(E_{kn'})-f(E_{kn}) \over E_{kn}-E_{kn'}}\,,
\end{align}
\end{subequations}
where $\chi_{\rm P}$ and $\chi_{\rm VV}$ denote the Pauli and the van Vleck contributions, respectively.
The spin susceptibility $\chi_s^{zz}$ (dashed lines) decreases to roughly half of its normal-state value (red dashed line). This residual value primarily arises from the van Vleck term $\chi_{\rm VV}$, 
{while the finite $\chi_{\rm P}$ is due to the residual $\rho(0)$. Interestingly, despite being a second-order transition, $\chi_P$ (blue solid line) exhibits behavior that appears discontinuous at $T_0$. In contrast, the total spin susceptibility $\chi_s^{zz}$ (blue dashed line) is continuous at $T_0$ and shows a conventional decrease below $T_0$. The unusual behavior of $\chi_P$ is attributed to energy splitting induced by spontaneous magnetic fields below $T_0$, which causes part of the Pauli term to shift to the van Vleck term. Indeed, when the external field becomes dominant over the spontaneous field, $\chi_P$ returns to conventional behavior. This peculiar behavior of $\chi_P$ just below $T_0$ serves as a hallmark of the chiral state with BFSs.~\cite{SM2} 
On the other hand, the jump in $\chi_s^{zz}$ at higher fields (gold-colored line) reflects the first-order nature of the transition.}


{\it Bogolon correlations}---
The presence of BFS suggests the possibility of additional instabilities in the superconducting phase, such as {Pomeranchuk or Cooper instabilities of bogolons}, both of which have been studied at zero field thus far \cite{PhysRevB.102.024505,PhysRevB.102.020501,PhysRevB.103.024521,PhysRevB.103.144517}.
Here, we explore the BFS instability in the $H-T$ phase diagram. 
We first consider {Pomeranchuk instabilities}, defined as
\begin{equation}
    \mathcal{D}_{\eta}=\sum_{\textbf{k}}\sum_n g_{\eta}(\textbf{k})\alpha_{\textbf{k}n}^{\dagger}\alpha_{\textbf{k}n},
\end{equation}
where $g_{\eta}(\textbf{k})$ is a form factor. For simplicity, we consider spherical harmonics $Y_{\ell m}(\hat{\textbf{k}})$ up to second order. 
{The corresponding bare correlation functions are given by}
\begin{equation}
    \bar\chi_{\eta}=\sum_{\textbf{k}}\sum_n |g_{\eta}(\textbf{k})|^2\left(-\frac{\partial f(E_{\textbf{k}n})}{\partial E_{\textbf{k}n}}\right).
\end{equation}
Figure\,3(a) displays the $T$ dependence of these correlations, which decrease monotonically with $T$, regardless of the form factor.
This behavior reflects the reduction in the DOS at the Fermi level caused by the opening of the superconducting gap. 
As depicted in Fig.\,3(b), the application of the Zeeman field partially restores the magnitude of these correlations; however, it does not exceed the normal-state value.  
This implies that Pomeranchuk instabilities are unlikely.

Next, we consider the bare correlation functions for the bogolon pairs, defined as
\begin{equation}
    \bar\phi_{\eta}=\sum_{\textbf{k}}\sum_n |g_{\eta}(\textbf{k})|^2\left(\frac{1-2f(E_{\textbf{k}n})}{2E_{\textbf{k}n}}\right).
\end{equation}
Figures\,3(c) and 3(d) show the temperature and field dependence of these correlations, respectively. Unlike {the case of the Pomeranchuk instability,} the bogolon pair correlation exhibits a logarithmic-like increase as the temperature decreases, due to the presence of BFS. This behavior indicates a potential transition into bogolon pairing states. 
{In the chiral spin-quintet state, the quasiparticle energies are non-degenerate due to the spontaneous magnetic field, allowing only odd-parity states in the absence of center-of-mass momentum. Here, let us consider $p$-wave and $f$-wave states, for simplicity.}
Among the $p$-wave pairing channels, the $p_x/p_y$-wave correlations dominate. These states are degenerate, and according to the BCS weak-coupling theory, a chiral $p_x+ip_y$-wave bogolon pairing state is expected to emerge at low temperatures under moderate interactions. In this state, the torus-shaped BFS near the $k_z=0$ plane becomes gapped, while the BFS near the $k_z$-axis remains intact.
Under high magnetic fields, the $p_z$-wave correlation becomes relatively enhanced {and overtakes the $p_x+ip_y$-wave correlation at the highest fields.} Figures 3(e) and 3(f) display contour maps of the correlation functions corresponding to the phase diagram of bogolon pairing states. These results suggest that a transition from the $p_x+ip_y$-wave state to the $p_z$-wave state is expected at low temperatures and under high fields. {This transition is accompanied by the disappearance of the BFS near the $k_z$-axis and the recovery of the torus-shaped BFS.

The next possible pairing channel is $f$-wave.
Similar to the $p$-wave case, the state with the highest $m=\ell$ becomes dominant at $H=0$, while states with lower $m$ are enhanced under high fields.~\cite{SM3}
The resulting contour maps are nearly identical to those observed in the $p$-wave case.
As shown in Ref.\,\onlinecite{PhysRevB.102.024505}, $p$-wave bogolon pairs primarily consist of $p$- and $f$-wave electron (hole) pairs. Therefore, if the usual electron-pairing interaction is sufficiently strong for the $p$- and $f$-wave channels, then $p\,(f)$-wave bogolon pairs may emerge.}

Furthermore, when these bogolon pairing states emerge, the hybridization of pairs with different symmetries, such as the $p$-wave and $d$-wave states, leads to spontaneous inversion symmetry breaking.~\cite{PhysRevB.102.020501,PhysRevB.103.024521,PhysRevB.103.144517} Thus, the additional transition observed at low temperatures and high fields, involving spontaneous inversion symmetry breaking, could provide supporting evidence for the BFS state.


{\it Conclusion}---
In this paper, we have theoretically investigated the evolution and instability of BFS in the spherical $j=3/2$ model under the Zeeman field. 
The application of the Zeeman field leads to significant growth of the BFS, particularly for those with a large weight of the $j_z=\pm 3/2$ component. Such evolution of the BFS can be observed by spectroscopic methods such as ARPES. 
Furthermore, the Pauli susceptibility shows a discontinuous-like behavior just below the transition temperature under zero field due to spontaneous magnetization. 
In addition, from our calculations of bogolon correlation functions, if an additional phase transition occurs in the superconducting phase, it is likely to be a chiral $p$- or $f$-wave bogolon pairing instability rather than Pomeranchuk instability. This chiral $p\,(f)$-wave bogolon pairing state, coexisting with the chiral $d$-wave superconducting state,  involves spontaneous inversion symmetry breaking. In this state, the torus-shaped BFS near the $k_z=0$ plane is gapped, while the BFS near the $k_z$ axis remains intact. Investigating the effects of magnetic fields on materials with BFS presents an interesting challenge for future research.


{\it Acknowledgments}---
{\footnotesize
We are grateful to S. Hoshino and T. Miki for their useful comments. This work was supported by KAKENHI Grants 
No.~19H01842, No.~19H05825, No.~23K25827, No.~24K01333  and JST SPRING, No.~JPMJSP2101.
}


\begin{thebibliography}{99}
\bibitem{Zeeman} {
The presence of the Zeeman potential proportional to $\hat{J}_z^3H^3$ further enhances the field effect on $j_z=\pm 3/2$ components.~\cite{PhysRev.102.1030,doi:10.7566/JPSJ.92.054703}  Therefore, considering this term will make the field-induced changes in the BFS more pronounced. However, we here focus on the conventional first-order Zeeman term only.}
\end{thebibliography}
\end{document}


\maketitle
\thispagestyle{plain}
\pagestyle{plain}

\section{$j_z$ weight on Fermi surface}
Figure S1 illustrates the first quadrant of the $k_x=0$ cut plane of two Fermi surfaces (FSs) in the normal state.
The blue and red colors indicate the weight of the $j_z=3/2$ and $1/2$ components on the Fermi surface, respectively.
We can see that the $j_z = 3/2$ component is prominent near the $k_z$ axis on the outer FS and near the $k_z = 0$ plane on the inner FS.

\begin{figure}[H]
\centering
\includegraphics[width=0.3\textwidth]{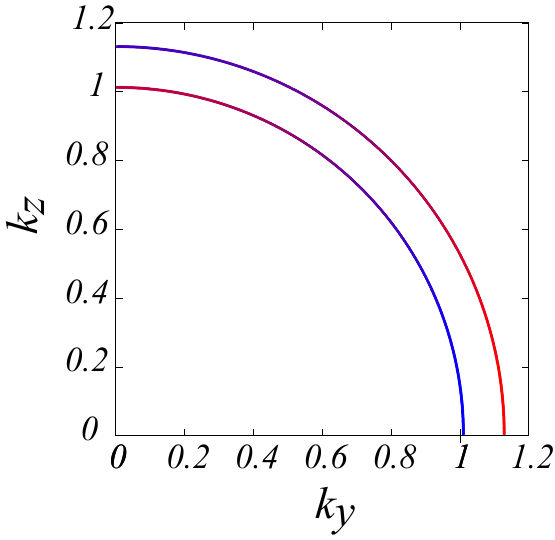}
\caption{$j_z$ weight on FS. The same one as Fig.\,2(a) in Ref.\,[1]. 
The first quadrant of the $k_x=0$ cut plane of the normal-state FSs.
The blue and red colors indicate the weight of the $j_z=3/2$ and $1/2$ components on the FSs, respectively.}
\end{figure}

\section{The quintet state and the conventional chiral $d$-wave state}
Figures S2(a) and S2(b) illustrate the $T$ dependence of the spin susceptibility at $H=0$ and the DOS at the Fermi level, respectively. Figure S2(c) depicts the DOS as a function of $E$. For comparison, the results for the conventional chiral $d$-wave state are also included.
The gap amplitude was not calculated self-consistently; instead, the values obtained in the quintet state were used.
As shown in Fig.\,S2(c), the conventional chiral $d$-wave state exhibits a characteristic $V$-shaped DOS, whereas the spin-quintet state shows a finite residual DOS near the Fermi level, corresponding to the presence of BFS.\,$^{2)}$
In the quintet state (blue line), the residual DOS increases as the spontaneous magnetization develops [Fig.\,2(b)] and exhibits a finite $\chi_{\rm P}$ at $T=0$. [Fig.\,S2(a)]. In contrast, no residual DOS is observed in the conventional chiral $d$-wave state (green line), and $\chi_{\rm P}$ is zero at $T=0$.
In Fig.\,S2(a), $\chi_P$ (blue solid line) shows discontinuous-like behavior at $T=T_0$ due to spontaneous magnetization.
In contrast, no such behavior is observed in the conventional chiral $d$-wave state (green solid line), because spontaneous magnetization is not induced within the present calculations.

\begin{figure}[t]
\centering
\includegraphics[width=\textwidth]{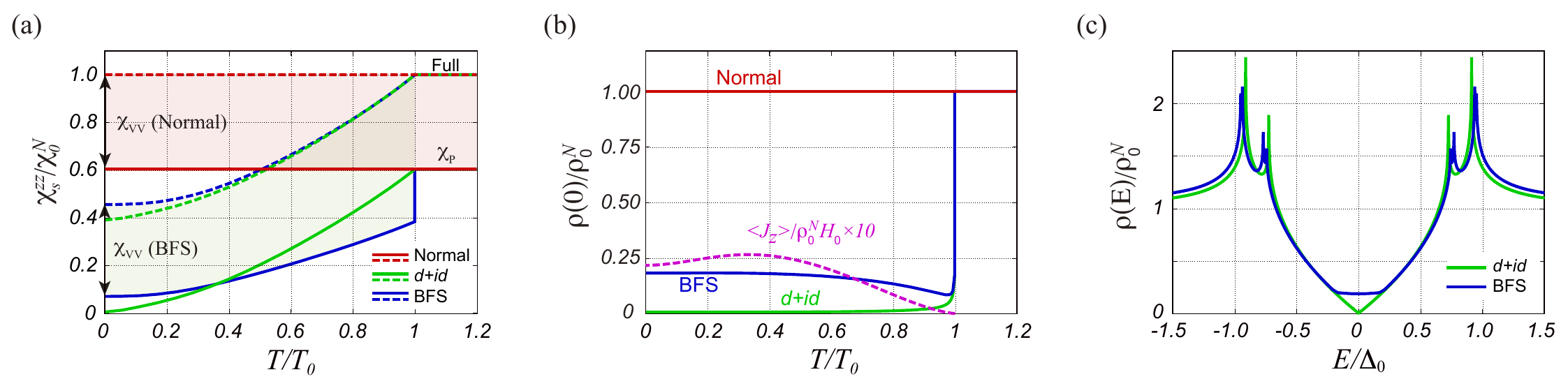}
\caption{(a) Spin susceptibility $\chi_s^{zz}$ normalized by the normal-state value $\chi_0^N$ as a function of $T$. 
The Pauli term $\chi_P$ and the van Vleck term $\chi_{VV}$ are denoted with solid and dashed lines, respectively.
$\chi_P$ in the quintet state (blue solid line) exhibits a discontinuous-like behavior at $T=T_0$, different from the conventional chiral $d$-wave state. 
The finite $\chi_P$ at $T=0$ corresponds to the residual DOS shown in panels (b) and (c).
(b) The DOS at the Fermi level and spontaneous magnetization. The residual DOS increases with the growth of spontaneous magnetization. 
(c) The DOS in the spin quintet state and the conventional chiral $d$-wave state as a function of $E$. 
}
\end{figure}


\section{Bogolon pair correlations}
Figures S3(a) and S3(b) illustrate the bogolon pair correlations for $\ell\le 21$ at $H=0$ and $H_0$, respectively, with the red lines corresponding to $\ell=21$.
The bogolon pair correlations are maximized at $m\sim\ell$ for $H=0$, but at $m\sim\ell/2$ for $H=H_0$. 
This is easily understood from the fact that the bogolon pair correlations can be expressed as
\begin{align}
\bar\phi_\eta=\sum_k |g_\eta(k)|^2\frac{\tanh(E_k/2T)}{2E_k}=4\pi\int_0^{\pi/2}\!\! h_\eta(\theta)\,\varphi_0(\theta)\sin\theta\,d\theta,
\end{align}
where $h_\eta(\theta)\equiv |g_\eta(k)|^2=|Y_{\ell m}|^2$ represents the squared basis functions, and
\begin{align}
\varphi_0(\theta)\equiv {1\over(2\pi)^3}\int k^2dk\,\frac{\tanh(E_k/2T)}{2E_k}.
\end{align}
\begin{figure*}[t]
\centering
\includegraphics[width=\textwidth]{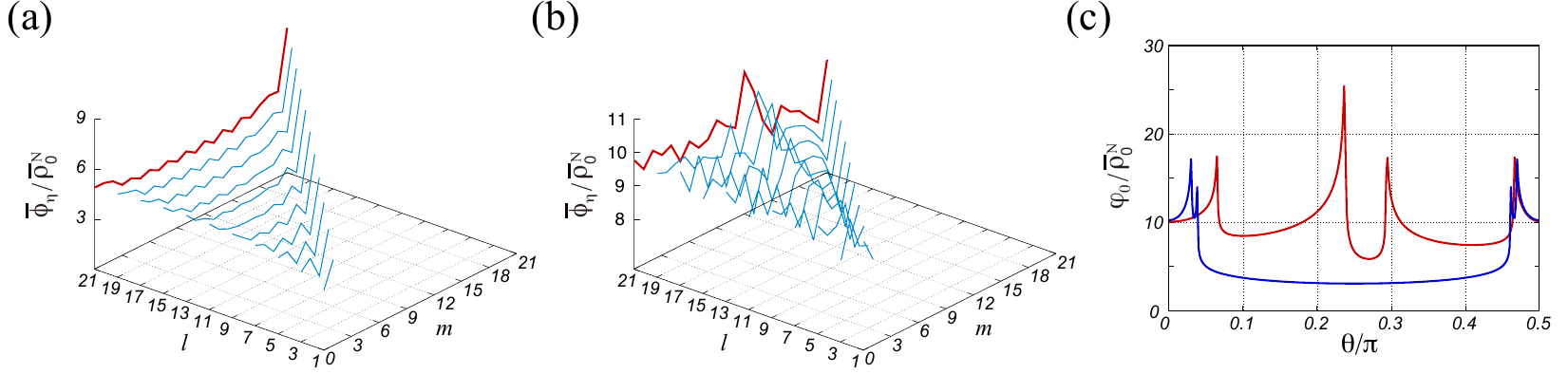}
\caption{Bogolon correlations $\bar\phi_\eta$ for $\ell\le 21$ at $H=0$ (a) and the highest field (b). The red lines correspond to $\ell=21$.
(c) The angular dependence of $\varphi_0(\theta)$ at $H=0$ (blue line) and $H_0$ (red line).}
\end{figure*}
Figure S3(c) shows the angular dependence of $\varphi_0(\theta)$ for $H=0$ (blue line) and $H_0$ (red line).
Since the bogolon pair correlations $\bar\phi_\eta$ are evaluated by multiplying $h_\eta(\theta)$ and $\varphi_0(\theta)\sin\theta$,
$\bar\phi_\eta$ is maximaized for $h_\eta$ that exhibits large value near $\theta\simeq\pi/2$ at $H=0$ ($\theta\simeq\pi/4$ at $H=H_0$). 

Although not shown here, $\bar\phi_\eta$ exhibits a similar structure for $\ell>21$, with its maximum at $m\sim\ell$ for $H=0$. 
The specific $(\ell,m)$ sets that exhibit strong correlations depend on the structure of $\varphi_0$, and consequently, on the size of the BFS.
However, the development of specific correlations strongly depends on the interaction, and such high-$\ell$ states are generally considered unlikely to emerge.
 

\vspace{20pt}

\noindent
{\bf References}\\
~~1)~T. Mori, H. Watanabe, and H. Ikeda, submitted to J. Phys.: Conf. Ser.\\
~~2)~H. Menke, C. Timm, and P. M. R. Brydon, Phys. Rev. B {\bf 100}, 224505 (2019).
